\documentclass[12pt]{article}

\usepackage[inline,shortlabels]{enumitem}
\usepackage{float}
\usepackage{mathtools}
\usepackage{subdepth}
\usepackage{comment}
\usepackage{graphicx}
\usepackage[round]{natbib}
\usepackage{bm}
\usepackage{algpseudocode}
\usepackage{listings}
\usepackage{tablefootnote}
\usepackage[margin=1in]{geometry}
\usepackage{tikz}
\usetikzlibrary{arrows.meta}
\usetikzlibrary{decorations.markings}
\usepackage[linesnumbered]{algorithm2e}
\RestyleAlgo{ruled}
\usetikzlibrary{plotmarks}
\usetikzlibrary{shapes,decorations,arrows,calc,arrows.meta,fit,positioning}
\usetikzlibrary{shapes.geometric}
\tikzset{
font={\fontsize{7pt}{12}\selectfont},
every node/.style={scale=1.3},
square/.style={regular polygon,regular polygon sides=4},
    -Latex,auto,node distance =1 cm and 1 cm,semithick,
    state/.style ={ellipse, draw, minimum width = 0.7 cm},
    block/.style = {square, draw, inner sep=0cm,minimum size=8mm},
    free/.style = {circle, draw, inner sep=0cm,minimum size=6mm},
    bidirected/.style={Latex-Latex,dashed},
    el/.style = {inner sep=2pt, align=left, sloped}
}\usepackage[english]{babel}
\usepackage{longtable}
\usepackage{nccmath}
\usepackage{color}
\usepackage{mathtools}
\usepackage{amssymb,amsmath,amsthm}
\usepackage{multirow,multicol,makecell,booktabs}
\usepackage[titletoc,title]{appendix}
\usepackage{authblk}
\usepackage{bbm}
\usepackage{setspace}
\usepackage{dsfont}
\usepackage[OT1]{fontenc}
\usepackage{subcaption}
\usepackage{refcount}
\usepackage{booktabs}
\graphicspath{ {../simulations/01_init_manuscript/graphs/} }
\usepackage{xr}
\usepackage[colorlinks,citecolor=blue,urlcolor=blue]{hyperref}

\AtEndDocument{\refstepcounter{theorem}\label{finalthm}}
\AtEndDocument{\refstepcounter{equation}\label{finaleq}}

{
  \theoremstyle{definition}
  \newtheorem{assumption}{}
}
{
  \theoremstyle{definition}
  
}
{
  \theoremstyle{definition}
  
}

\AtEndDocument{\refstepcounter{lemma}\label{finallemma}}

\renewcommand{\P}{\mathsf{P}}

\newcommand{\F}{\mathsf{F}}

\let\sec\S 

\renewcommand{\S}{\mathsf{S}}

\newcommand{\indep}{\mbox{$\perp\!\!\!\perp$}} 
 \newcommand{\dd}{\,\mathrm{d}}

\newcommand{\Zp}{Z^\pi}

\newcommand{\one}{\mathds{1}}
 
\newcommand{\E}{\mathsf{E}}
\renewcommand{\P}{\mathsf{P}}

\usepackage{accents}

\definecolor{codegreen}{rgb}{0,0.6,0}
\definecolor{codegray}{rgb}{0.5,0.5,0.5}
\definecolor{codepurple}{rgb}{0.58,0,0.82}
\definecolor{backcolour}{rgb}{0.95,0.95,0.92}

\lstdefinestyle{mystyle}{
    backgroundcolor=\color{backcolour},   
    commentstyle=\color{codegreen},
    keywordstyle=\color{magenta},
    numberstyle=\tiny\color{codegray},
    stringstyle=\color{codepurple},
    basicstyle=\ttfamily,
    showstringspaces=false,
    numbers=left
}

\DeclarePairedDelimiterX{\norm}[1]{\lVert}{\rVert}{#1}

\pgfdeclarelayer{background}
\pgfsetlayers{background,main}
\usetikzlibrary{arrows,positioning}
\tikzset{
>=stealth',
punkt/.style={
rectangle,
rounded corners,
draw=black, very thick,
text width=6.5em,
minimum height=2em,
text centered},
pil/.style={
->,
thick,
shorten <=2pt,
shorten >=2pt,}
}
\newcommand{\Vertex}[3]% pos, name
{\node[minimum width=0.6cm,inner sep=0.05cm] (#2) at (#1) {#3};
}
\newcommand{\Vertexr}[3]% pos, name
{\node[rectangle, draw, minimum width=0.6cm,inner sep=0.05cm] (#2) at (#1) {#2};
}
\newcommand{\ArrowR}[3]%
{ \begin{pgfonlayer}{background}
\draw[->,#3] (#1) to[bend right=30] (#2);
\end{pgfonlayer}
}
\newcommand{\ArrowLW}[3]%
{ \begin{pgfonlayer}{background}
\draw[->,#3] (#1) to[bend left=30] (#2);
\end{pgfonlayer}
}

\newcommand{\ArrowL}[3]%
{ \begin{pgfonlayer}{background}
    \draw[->,#3] (#1) to[bend left=45] (#2);
  \end{pgfonlayer}
}

\newcommand{\EdgeL}[3]%
{ \begin{pgfonlayer}{background}
\draw[dashed,#3] (#1) to[bend right=-45] (#2);
\end{pgfonlayer}
}

\newcommand{\Arrow}[3]%
{ \begin{pgfonlayer}{background}
\draw[->,#3] (#1) -- +(#2);
\end{pgfonlayer}
}

\allowdisplaybreaks
\pgfarrowsdeclare{arcs}{arcs}{...}
{
  \pgfsetdash{}{0pt} % do not dash
  \pgfsetroundjoin   % fix join
  \pgfsetroundcap    % fix cap
  \pgfpathmoveto{\pgfpoint{-5pt}{5pt}}
  \pgfpatharc{180}{270}{5pt}
  \pgfpatharc{90}{180}{5pt}
  \pgfusepathqstroke
}
\newcommand{\ArrowB}[3]%
{ \begin{pgfonlayer}{background}
    \draw[|-arcs,line width=0.4mm,shorten <= 0.3cm,shorten >= 0.3cm,#3] (#1) -- +(#2);
  \end{pgfonlayer}
}
\newcommand{\titlepaper}{\texttt{crumble}: A comprehensive framework for modern causal mediation analysis with intermediate confounding}

\date{\today}

\author[1,*]{Richard Liu}
\author[2]{Nicholas T. Williams}
\author[3]{Kara E. Rudolph}
\author[1]{Iv\'an D\'iaz}

\affil[1]{\small Division of Biostatistics, Department of Population
  Health, New York University Grossman School of Medicine.}

\affil[2]{\small Division of Biostatistics, University of California, Berkeley.}

\affil[3]{\small Department of Epidemiology, Mailman School of Public Health, Columbia University.}

\affil[*]{Corresponding Author. Email: \href{richard.l@nyu.edu}{richard.l@nyu.edu}}

\SetKwInput{KwInput}{Input}
\SetKwInput{KwOutput}{Output}
\SetKwInput{KwIf}{If}

\title{\titlepaper}
\begin{document}

%TC:ignore
\maketitle

\begin{abstract}
Causal mediation analysis is widely used to investigate how causal effects operate through specific pathways linking treatments or exposures to outcomes. Recently, \texttt{crumble} 
%\citep{williams2025crumble} %, the R package accompanying \citep{liu2024general}, 
was developed to enable nonparametric estimation of several mediation parameters, even when mediators are continuous and/or multi-dimensional or when treatments are non-binary. But a practical and accessible guide to using \texttt{crumble}---one that does not require deep familiarity with mediation analysis or semiparametric theory---is currently lacking. %Moreover, while still informative, existing tutorials on selecting appropriate mediation parameters predate recent theoretical developments, and tutorials incorporating these new insights remain scarce. 
This tutorial aims to %fill these gaps by providing a comprehensive yet 
an accessible introduction to \texttt{crumble} while minimizing technical complexity. We first review the mediation parameters implemented in \texttt{crumble}---natural direct and indirect effects, randomized interventional effects, and recanting-twin effects. For each, we give the definition, interpretation, identification assumptions, and suitability in the presence or absence of intermediate confounding. Then, we demonstrate the usage of \texttt{crumble} by examining an example configuration. Next, we describe how \texttt{crumble} accommodates non-binary treatments through modified treatment policies. Finally, we illustrate the practical use of \texttt{crumble} through two case studies---one with a binary treatment and one with a non-binary treatment---based on the Job Search Intervention Study data.
\end{abstract}
%TC:endignore
\newpage

%TC:ignore
\section{Introduction}
%TC:endignore

\label{intro}
This tutorial focuses on \texttt{crumble} \citep{williams2025crumble}, the R package that implements nonparametric estimators for causal mediation parameters \citep{liu2024general}. Causal mediation analysis evaluates the effects that treatments or exposures exert on outcomes through intermediate variables or mediators. Mediation analysis is widely applicable across many research fields. For example, clinicians may be interested in understanding the biological mechanisms by which vaccines (treatments) causally affect infection risk (outcome) \citep{benkeser2021inference}. In this context, immune responses can be chosen as mediators \citep{cowling2019influenza}, allowing researchers to quantitatively estimate the effects of vaccination on infection risk both through and not through specific immune responses.

There has been rapid development in the definition and identification of causal mediation parameters in recent decades, including but not limited to natural direct and indirect effects (NDE and NIE; \cite{robins1992,pearl2001direct}), randomized interventional direct and indirect effects (RIDE and RIIE; \cite{vanderLaan08, vanderweele2014effect, diaz2023efficient}), recanting-twin effects (RTEs; \citep{diaz2024non,vo2024recanting}), and mediation parameters based on separable effects \citep{robins2022interventionist,stensrud2021generalized} or stochastic interventions \citep{diaz2020causal,hejazi2023nonparametric}. This tutorial provides a comprehensive and user-friendly introduction to natural, randomized interventional, and recanting-twin effects. We note, however, that \texttt{crumble} can also be used to estimate the decision-theoretic approach to mediation analysis \citep{geneletti2007identifying} and organic direct and indirect effects \citep{lok2015organic,lok2016defining,lok2019causal,lok2021causal}.

NDE and NIE are among the most widely used mediation parameters with a mechanistic interpretation \citep{robins1992,pearl2001direct}. Nevertheless, their identification suffers from empirically untestable \textit{cross-world} counterfactual assumptions, which are violated when there are variables caused by the treatment/exposure that are common causes of the mediator and the outcome \citep{vanderweele2014effect}. Such variables, often referred to as \textit{intermediate confounders} or \textit{post-treatment confounders}, are common in scientific research. Thus, in the presence of such variables, researchers often consider alternative mediation parameters, such as RIDE, RIIE and RTE, which remain identifiable (under assumptions) even when intermediate confounders exist. 

While we acknowledge that important progress has been made in the definition and identification of mediation parameters, two significant factors limit practitioners from applying these parameters to real-world data. First, applied researchers often need guidance on which mediation parameters are most suitable for their scientific questions. While several tutorials provided such guidance (e.g., \cite{nguyen2021clarifying,rudolph2019causal}), they do not incorporate the more recently developed recanting twins estimators. Second, several challenges in \textit{non-parametric estimation} limit practitioners from applying these parameters to real-world data when the mediators are continuous and/or high-dimensional. Although some nonparametric estimators exist %First, to our best knowledge, few or no non-parametric methodologies were proposed for causal mediation analyses 
(e.g. \cite{rudolph2024practical}), these estimators exhibited stability challenges in finite samples. %However, continuous and high-dimensional mediators are common in many scientific domains. For example, epidemiologists may be interested in the causal effects of fat intake (exposure) on BMI (outcome) through the gut microbiome (mediator) \citep{bray1998dietary,sohn2017compositional}, where microbiome data are known to be continuous and high-dimensional. Second, most mediation parameters have been developed only for binary exposures, limiting their applicability when exposures are continuous or multivariate. Third, aside from \texttt{crumble} \citep{liu2024general,williams2025crumble}, no open-source R package provides non-parametric estimation of mediation parameters with continuous and/or high-dimensional mediators. Existing packages, such as \texttt{medoutcon} \citep{Hejazi2022}, \texttt{CMAverse} \citep{shi2021cmaverse}, and \texttt{mediation} \citep{tingley2014mediation}, either cannot handle complex mediators or focus solely on parametric estimation, which is prone to model misspecification and beyond the scope of this tutorial. 
%Although \citep{liu2024general} addressed these challenges with new statistical methodologies, it does not provide a comprehensive guide for using its accompanying R package 

In this tutorial, we introduce \texttt{crumble}, which harnesses the Riesz representation parameterization for improved estimator stability in the presence of continuous or multivariate mediators. %for practical mediation analysis. Such a guide would be particularly valuable for audiences who are not familiar with various types of mediation parameters or semi-parametric theory, including but not limited to epidemiologists, psychologists, biomedical scientists, and also statisticians in other subfields. This tutorial aims to provide a clear and comprehensive introduction to \texttt{crumble} while minimizing technical complexity. Specifically, 
We first review the definition, interpretation, and identification formulas of the five common mediation parameters, highlighting their advantages and limitations under different scenarios. Next, we demonstrate how to use \texttt{crumble} to conduct mediation analysis through one example configuration in the \texttt{crumble} website: \href{https://github.com/nt-williams/crumble/}{https://github.com/nt-williams/crumble/}. We conclude this tutorial with two case studies. 

\section{Notation and Set-up}\label{Notation}
Let $A \in \{0,1\}$ denote a binary treatment or exposure variable (the non-binary case is discussed in Section~\ref{nonbinary}), where $A = 0$ corresponds to the control or no-treatment group and $A = 1$ corresponds to the experimental or treatment group. Let $Z$ denote a vector of intermediate confounders (also referred to as post-treatment confounders), and note that NDE and NIE are not identifiable if such $Z$ exists. Throughout this tutorial, we use the terms ``treatment'' and ``exposure'' interchangeably, as well as ``intermediate confounder'' and ``post-treatment confounder'' interchangeably. Let $M$ denote a vector of mediators. Let $Y$ denote a continuous or binary outcome. Let $W$ denote a vector of covariates. Let $X = (W, A, Z, M, Y)$ denote a random variable with a true distribution $\P$. Let $X_1, \ldots, X_n$ be a sample of $n$ independent and identically distributed (i.i.d.) observations. %Let $\P f = \int f(x) \dd \P(x)$. Let $\P_n f = \frac 1n \sum_{i = 1}^n f(X_i)$. Let $\E_\F$ denote the expectation with respect to a distribution $\F$. When there is no ambiguity, we denote $\E = \E_\P$. 
We use the notation $A \sim B \mid C$ to indicate that $A$ and $B$ have the same conditional distribution given $C$, where $C$ may be empty. We use a hat to denote estimators. For example, we use $\E(Y \mid A = 1, W)$ to denote the true conditional expectation of $Y$ conditional on $(A =1, W)$, while we use $\hat \E(Y \mid A = 1, W)$ to denote an estimator of $\E(Y \mid A = 1, W)$.

%TC:ignore

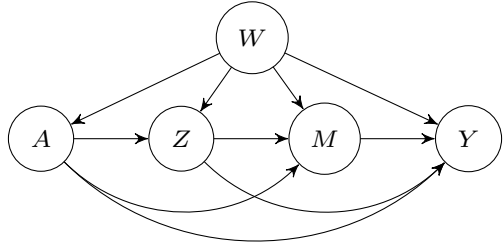
\begin{figure}[!htb]
  \centering
  \begin{tikzpicture}
    \tikzset{line width=1pt, outer sep=0pt,
      ell/.style={draw,fill=white, inner sep=2pt,
        line width=1pt}};
    \node[circle, draw, name=a, ]{$A$};
    \node[circle, draw, name=z, right = 10mm of a]{$Z$};
    \node[circle, draw, name=m, right = 10mm of z]{$M$};
    \node[circle, draw, name=y, right = 10mm of m]{$Y$};
    \node[circle, draw, name=w, above right = 7mm and 3mm of z]{$W$};

    \draw[->](a) to (z);
    \draw[->](w) to (z);
    \draw[->](z) to (m);
    \draw[->](a) to[out=-45,in=225] (y);
    \draw[->](a) to[out=-45,in=225] (m);
    \draw[->](z) to[out=-45,in=225] (y);
    \draw[->](w) to (y);
    \draw[->](m) to (y);
    \draw[->](w) to (a);
    \draw[->](w) to (m);
  \end{tikzpicture}
  \caption{The causal DAG. RIDE, RIIE and RTE are identifiable under this DAG; NDE and NIE are identifiable if Z does not exist ($Z = \emptyset$).}
  \label{fig:dag}  
\end{figure}

The directed acyclic graphs (DAGs) used across all five mediation parameters are given in Figure \ref{fig:dag}. 

%TC:endignore

The DAG implies the following structural causal model (SCM, \cite{Pearl00}): 

\begin{equation}
\begin{aligned}
& W = f_W(U_W); A = f_A(W, U_A); Z = f_{Z} (A, W, U_{Z}); \\
&M = f_{M} (A, W, Z, U_M); Y = f_Y (A, W, Z, M, U_Y),
\end{aligned}
\label{scm}
\end{equation}
where $(f_W, f_A, f_Z, f_M, f_Y)$ are fixed but unknown functions. $(U_W, U_A, U_Z, U_M, U_Y)$ are exogenous variables. For NDE and NIE where the DAG does not assume $Z$ exists, we modify (\ref{scm}) by removing the equation with respect to $Z$ and removing all $Z$s appearing in the input parameters of the functions. 

We also need to use the notation of counterfactual values in this tutorial. For any random variable $V$, under the SCM, we denote $V(a)$ as the counterfactual value of $V$ observed in a hypothetical world where $P(A = a) = 1$. Analogously, we denote $V(a,m)$ as that observed in a hypothetical world where $P(A = a, M = m) = 1$. For example, if $Z$ is present, $M(a) = f_M(a, W, Z(a), U_M)$ with $Z(a) = f_Z(a, W, U_Z)$; if $Z$ does not exist, $M(a) = f_M(a, W, U_M)$. 

\section{Review of Common Mediation Analysis Parameters}\label{ReviewParameter}
In this section, we review the definitions, interpretations, and identification of NDE, NIE, RIDE, RIIE, and RTE. For each parameter, we first introduce its \textit{causal estimand}, a quantity defined in terms of counterfactual variables that represents the target parameter of a causal question of interest. We then present the corresponding \textit{identification assumptions}, which link the causal estimand to a \textit{statistical estimand}. Finally, we describe the statistical estimand, typically expressed as a nested conditional expectation or an integral that can be estimated from observed data.

We note that previous tutorials have done this for natural and randomized interventional effects \citep{nguyen2021clarifying,rudolph2019causal}. However, our tutorial incorporates recent findings (e.g., \cite{miles2023causal}) regarding commonly used mediation parameters that were not previously recognized.
\subsection{Natural Direct and Indirect Effects}
\label{subsec:NDENIE}
Natural Direct and Indirect Effects (NDE and NIE) are defined as 
\begin{align}
    \operatorname{NIE} & = \E[Y(1, M(1)) - Y(1, M(0))], \label{NIE} \\ 
    \operatorname{NDE} & = \E[Y(1, M(0)) - Y(0, M(0))], \label{NDE}
\end{align}
where $\E[Y(a, M(a^*))]$ represents the average counterfactual outcome in a hypothetical world where all individuals receive treatment value $A = a$, while their mediator values are set to the natural value they would attain under $A = a^*$. NDE and NIE
decompose the average treatment effect (ATE), because $\operatorname{ATE} = \operatorname{NIE} + \operatorname{NDE}$. 

The NIE captures the average causal effect of altering the \textit{treatment value of the mediator} ($a^*$ in $\E[Y(a, M(a^*))]$, or the value of $A$ that enters the outcome directly in the SCM rather than through another variable same below), while holding the \textit{treatment value of the outcome} ($a$ in $\E[Y(a, M(a^*))]$, or the value of $A$ that enters the outcome indirectly in the SCM, same below) fixed at the level of ``treatment"\footnote{If the treatment value received by the outcome is fixed at the level of no-treatment, the resulting parameter is the \textit{pure direct effect} \citep{robins1992}. Although the methods discussed in this tutorial can, in principle, be applied to estimate pure direct and indirect effects, these quantities have not yet been implemented in \texttt{crumble} and are therefore outside the scope of this tutorial.}. Because the outcome’s treatment value remains the same in the two contrasting counterfactuals, any resulting change in $Y$ caused by intervening the treatment values must operate through the mediator $M$. Consequently, the NIE measures the effects through the causal pathway $A \to M \to Y$ in the left DAG of Figure~\ref{fig:dag}, which explains the term ``indirect effect.''

In contrast, the NDE measures the average causal effect of varying the treatment value of the outcome, while fixing the mediator to its natural value under no treatment. Since the mediator value is held fixed in the two contrasting counterfactuals, the causal effect from $A$ to $Y$ cannot pass through $M$. The NDE therefore captures the causal effect operating directly from $A$ to $Y$ (along the path $A \to Y$ in the left DAG of Figure~\ref{fig:dag}), hence the term “direct effect.”

While interpretations of the NIE/NDE that conceptualize interventions to ``set the mediator to the value it would have taken under treatment/control'' are sometimes useful, they may not be relevant for situations where the mediator is not a manipulable variable or interventions on the treatment value of the mediator are hard to conceptualize. In this case, the NIE/NDE (and the other three mediation parameters we will introduce below) may be interpreted simply as \textbf{describing variation in the data generating mechanisms}. Specifically, consider the SCM assumed in (\ref{scm}). Define the (random) function $g_{U}^N(a_1, a_2) = Y(a_1, M (a_2))$, then, we have 
\begin{align*}
    g_{U}^N(a_1, a_2)& = Y(a_1, M (a_2)) \\ &  = f_Y(W, a_1, Z(a_1), M(a_2), U_Y)  \quad \text{(by SCM)}\\
    & = f_Y(W, a_1, Z(a_1), M(a_2, Z(a_2)), U_Y). \quad \text{(counterfactual definition)}
\end{align*}
 Then, this bi-variate function tells us how the data generating mechanism varies as we vary the input parameters $a_1$ and $a_2$. The variation of $g_U$ along $a_1$ while holding $a_2$ fixed isolates to how the data generating mechanism for $Y$ responds to treatment through the pathways $A\to Y$ and $A\to Z\to Y$ (and only pathway $A \to Y$ if $Z$ does not exist), and the variation of $g_U$ along $a_2$ while holding $a_1$ fixed isolates to how the data generating mechanism for $Y$ responds to treatment through the pathways $A\to M \to Y$ and $A\to Z\to M \to Y$ (and only pathway $A \to M \to Y$ if $Z$ does not exist). The NDE is precisely a metric to measure how the function $g_U$ varies as a function of $a_1$ while holding $a_2$ fixed, and therefore is a metric to measure effects through paths not involving $M$. The NIE is a metric to measure how the function $g_U$ varies as a function of $a_2$ while holding $a_1$ fixed, and therefore is a metric to measure effects through paths involving $M$.

The assumptions for identifying NDE and NIE are as follows:

\begin{assumption}[Positivity]\label{assumptionpositivitynatural}
     \begin{enumerate}[label=(\roman*)]
        \item For all $w \in \operatorname{supp}(W)$, $\P(a \mid  w) > 0, a \in \{0, 1\}$,
        \item For all $a , a' \in \{0, 1 \}$, $\P( m \mid a, w)$ implies $\P( m \mid a', w)$,
        % \item For relevant $(a, m, w)$, $\P(Y \mid  a,  m,  w) > 0$ whenever $\P(m \mid  a, w ) > 0$.
    \end{enumerate}
\end{assumption}

\begin{assumption}[No unmeasured confounders]\label{assumption1} For all $(a,m)$:
\begin{enumerate}[label=(\roman*)]
    \item $A \indep Y(a, m) \mid W$,
    \item $M \indep Y(a, m) \mid W, A$,
    \item $A \indep M(a) \mid W$.
    \end{enumerate}
\end{assumption}

\begin{assumption}[Cross-world counterfactual independence]\label{assumption2} 
$M(0) \indep Y(1, m) \mid W$.
\end{assumption}

Assumption \ref{assumptionpositivitynatural} (and the analogous positivity assumptions required for other mediation parameters) ensures that the corresponding identification formulas are well defined. In practical terms, positivity requires that sufficient information is available for all variables of interest across every relevant subgroup of the data. That is, the data must contain meaningful variation so that the causal contrasts being defined can, in principle, be estimated. For example, part (i) of Assumption~\ref{assumptionpositivitynatural} requires that for every subgroup defined by the baseline covariates $W$, there are observed individuals in both the treatment and control groups. 

Assumption \ref{assumption1} holds if there is no unmeasured confounder that is a common cause of any pair of variables in $(A, M, Y)$. Assumption \ref{assumption2}  requires two counterfactual variables defined in different counterfactual worlds to be conditionally independent, which is neither testable by data nor enforceable by any study design. This has often been cited as a reason to avoid natural direct and indirect effects \citep{vansteelandt2017interventional}. The assumption \ref{assumption2} will be violated when an intermediate confounder $Z$ exists, which is affected by $A$ and is a common cause of $M$ and $Y$. For example, in a hypothetical randomized clinical trial, if clinicians believe that treatment assignment works through patients' adherence status, then the cross-world assumption will not hold as adherence status is a post-treatment confounder. Importantly, causal identification assumptions such as Assumption \ref{assumption1} are often assessed through the requirement of no unmeasured confounders \citep{vanderweele2019principles}. E.g., $M\indep Y(a,m)\mid (W, A)$ holds if $(A, W)$ contains all common causes of $M$ and $Y$. In terms of the structural model, this means that $U_M\indep U_Y\mid (A, W)$ implies $M\indep Y(a,m)\mid (W, A)$. The former assumption has a clear interpretation that can be communicated to subject matter experts: all the common causes of $M$ and $Y$ are in $(A,W)$, which means there must be no post-treatment common causes of the mediator and outcome. Importantly, this same structural assumption $U_M\indep U_Y\mid (A, W)$ implies the counterfactual independence $M(0) \indep Y(1, m) \mid W$. Therefore, cross-world counterfactual independence can be argued using subject-matter expert knowledge of the same kind that is commonly used to argue the assumption of no unmeasured confounders in practice. 

While the cross-world nature of the assumptions may not be problematic, the fact that it precludes the existence of intermediate confounders is much more so, as such confounders are common in research questions. This motivates the development of other mediation parameters that can be identified when intermediate confounders exist in the DAG, which we will discuss in \sec\ref{rie} and \sec\ref{rte}. 

Under assumptions \ref{assumptionpositivitynatural} to \ref{assumption2}, NDE and NIE are identified as 
\begin{align*}
    \operatorname{NIE}  = \psi^{natural}(1, 1) - \psi^{natural}(1, 0), \operatorname{NDE}  = \psi^{natural}(1, 0) - \psi^{natural}(0, 0),
\end{align*}
where 
\begin{align*}
    \psi^{natural}(a_1, a_2) & = \E [ \E \{ \E (Y \mid A = a_1, M, W)\mid A = a_2, W\}] \\ 
    & = \int_w \int _m \E(Y \mid a_1, m, w) \dd \P (m \mid a_2, w) \dd \P(w).
\end{align*}

\subsection{Randomized Interventional Direct and Indirect Effects} \label{rie}
Randomized Interventional Direct and Indirect Effects (RIDE and RIIE) are defined as 
\begin{align}
    \operatorname{RIIE} = \E[Y(1, G(1)) - Y(1, G(0))], \\ 
    \operatorname{RIDE} = \E[Y(1, G(0)) - Y(0, G(0))],
\end{align}
where $G(a) \sim M(a) \mid W$. In these definitions, $E[Y(a, G(a^*))]$ represents the average outcome in a hypothetical world where all individuals receive treatment $A = a$, while their mediator values are replaced by a random draw $G(a^*)$ that follows the same conditional distribution as $M(a^*)$ given $W$. 

The RIIE can be interpreted as the average difference between two counterfactual worlds where the treatment value of the outcome ($a$ in $Y(a, G(a^*))$) in both is set to be under treatment, while the mediator of the outcome ($G(a^*)$ in  $Y(a, G(a^*))$) was set to be the mediator's distribution under different treatment levels. RIIE was termed an ``indirect effect'' because, like the NIE, it contrasts two hypothetical worlds in which the mediator distribution varies while the treatment level is fixed, meaning that the effect caused by changing the treatment value must pass ``indirectly" through the mediator. Conversely, the RIDE is termed a ``direct effect'' because, analogous to the NDE, it contrasts two hypothetical worlds in which the treatment level varies while the mediator distribution is held fixed, meaning that the effect caused by changing the treatment value passes ``directly" to the outcome. The RIIE and RIDE can also be understood as a description of the variation in the data generating mechanisms. In detail, define $g_U^R(a_1, a_2) = Y(a_1, G(a_2)) = f_Y\bigl(W, a_1, Z(a_1), G(a_2), U_Y\bigr)$, then, under this representation, the RIIE corresponds to how the function $g_U^R$ varies as a function of $a_2$ while holding $a_1$ fixed, whereas the RIDE corresponds to how the function $g_U^R$ varies as a function of $a_1$ while holding $a_2$ fixed.

Randomized interventional effects (RIEs) are widely used in settings where the exposure is not a manipulable variable. For example, consider the racial health disparity setting discussed in \citep{vanderweele2017mediation}. Let $A$ denote race, where $A = 1$ represents Black individuals and $A = 0$ represents White individuals. Let $M$ denote socioeconomic status, and let $Y$ denote a health outcome. In this context, the RIIE can be interpreted as the remaining health disparity if the distribution of socioeconomic status among Black individuals, $G(1)$, were set to that of White individuals, $G(0)$. Similarly, the RIDE can be interpreted as the health disparity between Black and White individuals if the distribution of socioeconomic status were set to that of White individuals. 

While such interpretations are reasonable, it is important to note that RIEs are not the only parameters for mediation analysis when the exposure is not manipulable. For example, natural effects can also be used to study racial health disparities \citep{jackson2018decomposition,zhou2022semiparametric,diaz2024non,ou2025assessing} because they admit an alternative interpretation in terms of variation in the data-generating mechanisms, as discussed in Section \ref{subsec:NDENIE}. In other words, although the counterfactual query entailed by natrual effects ``what a Black individual’s socioeconomic status would have been had they been of a different race" (cf. \cite{vanderweele2017mediation}) sound strange, this is not the only interpretation available for natural effects.

% Related discussion on interpreting RIEs can be found in  \citep{vanderweele2014causal,jackson2021meaningful}. 

The assumptions for identification of RIDE and RIIE are listed as follows:

\begin{assumption}[Positivity]\label{assumptionpositivityrandomized}
\begin{enumerate}[label=(\roman*)]
    \item For all $w \in \operatorname{supp}(W)$, $\P( a \mid w) > 0, a \in \{0, 1\}$,
    \item For relevant $z$ and any $a, a^* \in \{0, 1\}$, $\P(m \mid a^*, w) > 0$ whenever $\P(m \mid a, z, w) > 0$,
    \item For relevant $(z, w, m)$ and any $a, a^* \in \{0, 1\}$, $\P( m \mid  a,  z, w) > 0$ whenever $\P(z \mid a, w) > 0$ and $\P(m \mid a^*, w) > 0$,
\end{enumerate}
\end{assumption}
\begin{assumption}[No unmeasured confounders]\label{assumption3} For all $(a,m)$:
\begin{enumerate}[label=(\roman*)]
    \item $A \indep Y(a, m) \mid W$,
    \item $M \indep Y(a, m) \mid W, Z, A$,
    \item $A \indep M(a) \mid W$.
    \end{enumerate}
\end{assumption}

Assumption \ref{assumption3} will hold if there is no unmeasured confounder for any pair of $(A, M, Y)$. Notably, cross-world assumptions are not needed for identifying randomized interventional effects, meaning that they can still be identified when $Z$ is present. While promising, two important drawbacks arise. First, a recent paper \citep{miles2023causal} identified a fundamental limitation of RIIE: It does not necessarily capture the true \textit{mechanistic} indirect effect. For example, if some units experience treatment-mediator effects, the other units experience mediator-outcome effects, and no unit experiences both, then any mechanistic indirect effect measure should be zero. The RIIE, however, can be nonzero in this scenario. To illustrate, consider the example in Section \ref{intro}, where $A$ denotes vaccine status, $M$ denotes immune response, $Y$ denotes infection risk, and $Z$ denotes an unobserved post-treatment confounder related to immune response (e.g. another immune response that is not of primary interest). Suppose, hypothetically, that $\hat{\text{RIIE}} = 0.5$. In general, this cannot be interpreted mechanistically as ``the infection risk increases by 0.5 units due to the change in immune response ($M$) induced by vaccination.'' In contrast, if $\hat{\text{NIE}} = 0.5$, such a mechanistic interpretation would be appropriate. The RIIE retains a meaningful non-mediational interpretation in certain settings, such as studies of racial health disparities; see Sections 6 and 8 of \cite{miles2023causal} for further discussion. Second, compared with natural effects, randomized interventional effects do not generally decompose the average treatment effect (ATE) in the presence of intermediate confounding by $Z$; that is, $\operatorname{RIDE} + \operatorname{RIIE} \neq \operatorname{ATE}$ when $Z$ is present (cf. Corollary 1 in \cite{miles2023causal}). Returning to the vaccine example, suppose there is intermediate confounding by $Z$ and that $\hat{\text{RIIE}} = 0.5$ and $\hat{\text{ATE}} = 1$. Then the interpretation that ``the indirect effect accounts for 50\% of the ATE'' is invalid.

Under Assumptions \ref{assumptionpositivityrandomized} and \ref{assumption3}, RIDE and RIIE are identified as 
\begin{align*}
    \operatorname{RIIE} = \psi ^{rand}(1, 1) - \psi ^{rand}(1, 0), \operatorname{RIDE} = \psi ^{rand}(1, 0) - \psi ^{rand}(0, 0),
\end{align*}
where 
\begin{align*}
    \psi^{rand} (a_1, a_2) & = \E \left[ \E \left\{ \int_z \E(Y \mid A = a_1, z, M,W) \dd \P( z \mid A = a_1, W) \mid A = a_2, W \right\} \right] \\ 
    & = \int_w\int_m \int_z\E(Y \mid A = a_1, z, m,w) \dd \P( z \mid A = a_1, w) \dd \P (m \mid A= a_2, w) \dd \P (w)
\end{align*}

\subsection{Recanting-twin Effects} \label{rte}
Recanting-twin Effects (RTEs, \cite{diaz2024non,vo2024recanting}) are concerned with \textit{path-specific} effects from $A$ to $Y$. Define $P_1: A\rightarrow Y$;
$P_2: A \rightarrow Z \rightarrow Y$;
$P_3: A \rightarrow Z \rightarrow M \rightarrow Y$ and
$P_4: A \rightarrow M \rightarrow Y$, and the
following nested counterfactuals:

\begin{align}
  Y_{S_0}&=Y(1, Z(1), M(1, Z(1))),\notag\\
  Y_{S_1}&=Y(0, Z(1), M(1, Z(1))),\notag\\
  Y_{S_2}&=Y(0, Z(0), M(1, Z(1))),\label{eq:count} \\
  Y_{S_3}&=Y(0, Z(0), M(1, Z(0))),\notag\\
  Y_{S_4}&=Y(0, Z(0), M(0, Z(0))),\notag
\end{align}
then the natural path-specific effects \citep{pearl2001direct} through $P_j$ could be defined as $E(Y_{S_{j - 1}} - Y_{S_j})$. To make this result more transparent, we define $g_U^N(a_1, a_2, a_3, a_4) = Y(a_1, Z(a_2), M(a_3, Z(a_4)))$, then $E(Y_{S_{j - 1}} - Y_{S_j})$ can be understood as the average difference of varying $a_j$ while keeping all other $a_k$s ($k \ne j$) fixed, which is exactly the effects only through path $P_j$ but not others. Additionally, note that $\sum_{j = 1}^4 \E(Y_{S_{j - 1}} - Y_{S_j}) = E(Y(1) - Y(0)) = \operatorname{ATE}$, meaning that ATE can be decomposed as the natural path-specific effects through $\{P_j\}_{j = 1}^4$. 

Unfortunately, the natural path-specific effects for $P_2$ and $P_3$ are not identifiable due to the non-identifiability of the distribution of $Y_{S_2}$. Identifying the distribution  of $Y_{S_2}$ would require identification of the joint distribution of $(Z(1), Z(0))$, which is not identifiable in any SCM without imposing strong assumptions \citep{tchetgen2014identification}. Intuitively, identifying this joint distribution from observed data would require observing the same individual under both treatment and no-treatment conditions. Under the DAG in Figure \ref{fig:dag}, such a joint observation is impossible, as one individual can only be under treatment or under no-treatment. The recanting twin effects of $P_2$ and $P_3$ aim to resolve this non-identifiability issue caused by the joint distribution of $(Z(1), Z(0))$. In general, the idea is to substitute either $Z(1)$ or $Z(0)$ into a ``random draw" following the same conditional distribution, so that the troublesome joint distribution of $(Z(1), Z(0))$ can be transformed into identifiable probability distributions. Specifically, for $P_2$ and $P_3$, a random draw $T(a) \sim Z(a) \mid W$, $a \in \{0,1\}$, was introduced to replace one of the $Z(a)$s in $Y_{S_2}$. This substitution transforms the non-identifiable joint distribution of $(Z(1), Z(0))$ into a joint distribution of either $(Z(1), T(0))$ or $(T(1), Z(0))$, which is identifiable because of the assumed conditional independence between $T(a)$ and $Z(a')$ for any $a, a' \in \{0, 1\}^2$. For example, because 
\begin{align*}
    \P(Z(1), T(0) \mid W) & = \P(Z(1) \mid W)\P(T(0) \mid W) \quad \text{(conditional independence)}\\ 
    & = \P(Z(1) \mid W)\P(Z(0) \mid W) \quad (T(0) \sim Z(0) \mid W),
\end{align*}
we know that $\P(Z(1), T(0) \mid W)$ is identifiable because either $\P(Z(1) \mid W)$ or $\P(Z(0) \mid W)$ is identifiable.

The RTEs for $P_2$ and $P_3$ are based on the following modified nested counterfactuals:

\begin{align*}
  Y_{S_1}' &= Y(0, Z(1), M(1, T(1))),\qquad
  Y_{S_2}'' = Y(0, T(0), M(1, Z(1))),\\
  Y_{S_2}' &= Y(0, Z(0), M(1, T(1))),\qquad
  Y_{S_3}'' = Y(0, T(0), M(1, Z(0))).
\end{align*}
With these modified nested counterfactuals, the RTEs through $P_1$ to $P_4$ were defined as
\begin{align} \label{eq:P1toP4}
   \psi_{P_1} = \E(Y_{S_0} - Y_{S_1}), \quad \psi_{P_2} = \E(Y_{S_1}' - Y_{S_2}'),\quad
  \psi_{P_3} = \E(Y_{S_2}'' - Y_{S_3}''), \quad \psi_{P_4} = \E(Y_{S_3} - Y_{S_4}).
\end{align}
Using RTEs, the ATE can be alternatively decomposed as 
\begin{align*}
    \operatorname{ATE} = \sum_{j = 1}^4 \E(Y_{S_{j - 1}} - Y_{S_j}) = \psi_{P_1} + \psi_{P_2} + \psi_{P_3} + \psi_{P_4} + R,
\end{align*}
where 
\begin{equation} \label{eq:R}
    R = \E (Y_{S_1} - Y_{S_1}' + Y_{S_2}' - Y_{S_2}'' +Y_{S_3}'' - Y_{S_3})
\end{equation}
is a remainder term generated by substituting natural path-specific effects with RTEs. 

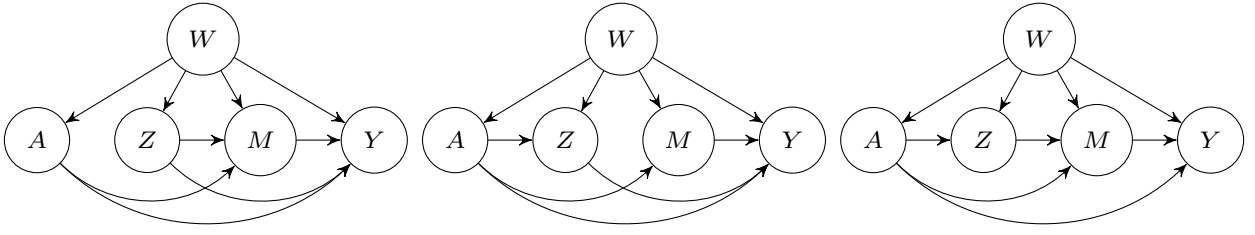
\begin{figure}[!htb]
  \centering
  \begin{tikzpicture}
    \tikzset{line width=1pt, outer sep=0pt,
      ell/.style={draw,fill=white, inner sep=1pt,
        line width=1pt}};
    \node[circle, draw, name=a, ]{$A$};
    \node[circle, draw, name=z, right = 6mm of a]{$Z$};
    \node[circle, draw, name=m, right = 6mm of z]{$M$};
    \node[circle, draw, name=y, right = 6mm of m]{$Y$};
    \node[circle, draw, name=w, above right = 7mm and 1mm of z]{$W$};

    % \draw[->](a) to (z);
    \draw[->](w) to (z);
    \draw[->](z) to (m);
    \draw[->](a) to[out=-45,in=225] (y);
    \draw[->](a) to[out=-45,in=225] (m);
    \draw[->](z) to[out=-45,in=225] (y);
    \draw[->](w) to (y);
    \draw[->](m) to (y);
    \draw[->](w) to (a);
    \draw[->](w) to (m);
  \end{tikzpicture}
  \hfill
  \begin{tikzpicture}
    \tikzset{line width=1pt, outer sep=0pt,
      ell/.style={draw,fill=white, inner sep=2pt,
        line width=1pt}};
    \node[circle, draw, name=a, ]{$A$};
    \node[circle, draw, name=z, right = 6mm of a]{$Z$};
    \node[circle, draw, name=m, right = 6mm of z]{$M$};
    \node[circle, draw, name=y, right = 6mm of m]{$Y$};
    \node[circle, draw, name=w, above right = 7mm and 1mm of z]{$W$};

    \draw[->](a) to (z);
    \draw[->](w) to (z);
    % \draw[->](z) to (m);
    \draw[->](a) to[out=-45,in=225] (y);
    \draw[->](a) to[out=-45,in=225] (m);
    \draw[->](z) to[out=-45,in=225] (y);
    \draw[->](w) to (y);
    \draw[->](m) to (y);
    \draw[->](w) to (a);
    \draw[->](w) to (m);
  \end{tikzpicture}
  \hfill
  \begin{tikzpicture}
    \tikzset{line width=1pt, outer sep=0pt,
      ell/.style={draw,fill=white, inner sep=2pt,
        line width=1pt}};
    \node[circle, draw, name=a, ]{$A$};
    \node[circle, draw, name=z, right = 6mm of a]{$Z$};
    \node[circle, draw, name=m, right = 6mm of z]{$M$};
    \node[circle, draw, name=y, right = 6mm of m]{$Y$};
    \node[circle, draw, name=w, above right = 7mm and 1mm of z]{$W$};

    \draw[->](a) to (z);
    \draw[->](w) to (z);
    \draw[->](z) to (m);
    \draw[->](a) to[out=-45,in=225] (y);
    \draw[->](a) to[out=-45,in=225] (m);
    % \draw[->](z) to[out=-45,in=225] (y);
    \draw[->](w) to (y);
    \draw[->](m) to (y);
    \draw[->](w) to (a);
    \draw[->](w) to (m);
  \end{tikzpicture}
  \caption{Three DAGs with ``no intermediate confounding" due to the fact that in either of the three scenarios, $Z$ is no longer a valid intermediate confounder/post-treatment confounder. \textbf{Left}: The scenario where $A \not \to Z$, meaning that in the population, $A$ will not cause $Z$; \textbf{Middle}: The scenario where $Z \not \to M$; \textbf{Right}: The scenario where $Z \not \to Y$. The concept of intermediate confounding was formally defined in Definition 5 of \citep{vo2024recanting}.}
  \label{fig:daginterconfounding}  
\end{figure}

We note that except for being identifiable when $Z$ exists in the DAG, RTEs enjoy two further desirable properties. First, similar to the NDE and NIE, $\psi_{P_i}, i =1, 2, 3, 4$ measures the true mechanism of the corresponding path. That means, if no unit in the population experiences a causal effect through path $P_i$, then $\psi_{P_i} = 0$ (see technical details in Section 5 of \cite{diaz2024non}). Second, according to Lemma 1 in \citep{vo2024recanting}, if there is no intermediate confounding by $Z$ (see Figure \ref{fig:daginterconfounding} for a visual illustration), then $R = 0$. Therefore, a falsification test on the null hypothesis $H_0: R = 0$ can be conducted. If the test rejects the null, then there is evidence of intermediate confounding by $Z$. In that case, omitting $Z$ from the analysis and proceeding to estimate the NDE or NIE would generally yield unreliable estimates, because the identification assumptions underlying these effects would not hold. Conversely, if the test fails to reject the null and substantive knowledge also suggests that $Z$ is unlikely to be an intermediate confounder, then interpreting RTEs as natural path-specific effects may be appropriate.

%TC:ignore

%TC:endignore
The identification assumptions for RTEs are listed as follows:
\begin{assumption}[Positivity]\label{assumptionpositivityrecantingtwin}
All densities used in the identification formula are strictly positive. 
\end{assumption}
\begin{assumption}[Sequential ignorability]\label{assumption4} For all $(a,m,z)$:
\begin{enumerate}[label=(\roman*)]
    \item $Y(a, m, z) \indep A \mid W;$ $Y(a, m, z) \indep Z \mid (A = a, W);$ and $Y(a, m, z) \indep M \mid (A = a, Z = z, W)$,
    \item $M(a, z) \indep A \mid W;$ and $M(a, z) \indep Z \mid (A = a, W)$
    \item $(Z(a), M(a)) \indep A \mid W$.
    \end{enumerate}
\end{assumption}

\begin{assumption}[Cross-world counterfactual independence]\label{assumption5} For all $a, a', a'' = 0, 1$; $z, z' \in \operatorname{supp}(Z)$ and $m \in \operatorname{supp}(M)$:
\begin{enumerate}[label=(\roman*)]
    \item $Y(a, m, z) \indep (M(a', z'), Z(a'')) \mid W ;$
    \item $M(a, z) \indep Z(a') \mid W$.
    \end{enumerate}
\end{assumption}

As in the case of the NDE/NIE, Assumptions \ref{assumption4} and \ref{assumption5} will hold if a researcher measures all common causes of any pair of $(A, Z, M, Y)$. In other words, different from NDE and NIE, there is no need for researchers to reason about whether \ref{assumption5} is testable: as long as they can argue that all common causes of the relevant variables are measured, these assumptions should be satisfied. We refer the reader to \cite{vo2024recanting} for more discussion on this point. 

The identification results under assumptions \ref{assumptionpositivityrecantingtwin} to \ref{assumption5} for $E(Y_{S_0})$, $E(Y_{S_1})$, $E(Y_{S_1}')$, $E(Y_{S_2}')$, $E(Y_{S_2}'')$, $E(Y_{S_3})$, $E(Y_{S_3}'')$, and $E(Y_{S_4})$ are given in Equation (6) in \citep{vo2024recanting}. Then, one can obtain $\psi_{P_j}, j = 1, 2, 3, 4$ and $R$ by equations (\ref{eq:P1toP4}) and (\ref{eq:R}).
% {\small 
% \begin{align}
%     \E(Y_{S_0}) &= \int_w \E(Y\mid A = 1, w) \dd \P (w), \notag\\
%         \E(Y_{S_1}) &=\int_w \int_m \int_z\E(Y\mid A = 0, z, m, w)\dd \P(m,z\mid A = 1, w) \dd \P (w), \notag\\
%     \E(Y_{S_1}') &= \int_w \int_m \int_z \E(Y\mid A = 0,z,m,W)\dd \P (m\mid A = 1, W) \dd \P(z\mid A = 1, w) \dd \P(w), \notag\\
%       \E(Y_{S_2}')  &= \int_w \int_m \int_z \E(Y\mid A = 0,z,m,w)\dd \P (m\mid A = 1, w) \dd \P(z\mid A = 0, w) \dd \P (w), \label{eq:iden}\\
%       \E(Y_{S_2}'') &= \E(Y_{S_2}'), \notag\\
%     \E(Y_{S_3})  &= \int_w \int_m \int_z \E(Y\mid A = 0,z,m,w)\dd \P (m\mid A = 1, z,w) \dd \P(z\mid A = 0, w) \dd \P(w), \notag\\
%     \E(Y_{S_3}'')  &= \int_w \int_m \int_z \int_{z'} \E(Y\mid A = 0,z,m,W) \dd \P (m\mid A = 1, z',W) \dd \P(z\mid A = 0, W) \dd \P(z'\mid A = 0, W)\dd \P (w), \notag\\
%     \E(Y_{S_4}) &= \int_w \E(Y\mid A = 0, w) \dd \P (w), \notag
% \end{align}}

% \section{Estimation: Challenges and Solutions in \texttt{crumble}}
\section{Demonstration of \texttt{crumble} usage}
\label{sec:demo}

In this section, we illustrate the use of \texttt{crumble} by explaining a specific example provided at the README page of the \texttt{crumble} Github: \texttt{\href{https://github.com/nt-williams/crumble/}{https://github.com/nt-williams/crumble/}}. In some informal communications (conversations/emails), we also found that applied researchers showed great interest in connecting the software design with technical details in \citep{liu2024general}. For this purpose, we provided a more technical introduction of the estimation methodology in Appendix C.

Mediation analysis can be readily conducted using the \texttt{crumble()} function. By specifying the required input parameters (which encode the necessary components of the mediation analysis), users can directly obtain interval estimates of the mediation parameter of interest. An example from the GitHub repository is shown below.

\begin{verbatim}
    crumble(
        data = weight_behavior,
        trt = "sports", 
        outcome = "bmi",
        covar = c("age", "sex", "tvhours"),
        mediators = c("exercises", "overweigh"),
        moc = "snack",
        d0 = \(data, trt) factor(rep(1, nrow(data)), levels = c("1", "2")), 
        d1 = \(data, trt) factor(rep(2, nrow(data)), levels = c("1", "2")), 
        effect = "RT",
        learners = c("mean", "glm", "earth", "ranger"), 
        nn_module = sequential_module(),
        control = crumble_control(crossfit_folds = 1L, epochs = 20L)
)
\end{verbatim}

We explain the configuration of this example line by line. The parameter \texttt{data} (line 2) must be a tabular dataset in which each row corresponds to an individual with observed variables $(W_i, A_i, Z_i, M_i, Y_i)$. All variables must be numeric. In this example, \texttt{weight\_behavior} is a dataset from the \texttt{mma} package \citep{yu2017mma}. Rows with missing values have been removed, as \texttt{crumble()} cannot be applied to datasets with missingness.

Lines 3 to 7 specify the roles of variables (i.e., column names) in the dataset. Specifically, \texttt{trt} (line 3) defines the treatment variable $A$; in this example, \texttt{sports} serves as the treatment, \texttt{outcome} (line 4) specifies the outcome variable $Y$, \texttt{covar} (line 5) specifies the covariates $W$, \texttt{mediators} (line 6) specifies the mediator(s) $M$, and \texttt{moc} (line 7) specifies the intermediate confounder $Z$ (where \texttt{moc} stands for ``mediator--outcome confounder'').

The parameters \texttt{d0} (line 8) and \texttt{d1} (line 9) specify the treatment hypothetical interventions, where \texttt{d0} and \texttt{d1} correspond to the reference treatment and active treatment hypothetical interventions, respectively. In this example, \texttt{d0} represents the intervention in which all individuals are assigned $A = 1$, while \texttt{d1} represents the intervention in which all individuals are assigned $A = 2$. This reflects the coding in the dataset, where $A = 1$ denotes no-treatment and $A = 2$ denotes treatment.

The parameter \texttt{effect} (line 10) specifies the mediation parameter of interest. In this example, \texttt{RT} corresponds to RTEs. Users may alternatively specify \texttt{effect} as \texttt{N} (natural direct and indirect effects, NDE/NIE), \texttt{RI} (randomized interventional direct and indirect effects, RIDE/RIIE), or \texttt{O} (organic effects, which are not covered in this tutorial).

The parameter \texttt{learners} (line 11) specifies the set of candidate algorithms used by \texttt{mlr3superlearner}, an \texttt{R} package that implements the Super Learner \citep{vanderLaanPolleyHubbard07}. The Super Learner is an ensemble method employed throughout the estimation process and requires the user to provide a library of candidate algorithms. The options for \texttt{learners} must be selected from those implemented in \texttt{mlr3superlearner}, which are available at \href{https://github.com/nt-williams/mlr3superlearner}{https://github.com/nt-williams/mlr3superlearner}. In practice, to ensure that the Super Learner achieves satisfactory predictive performance, we recommend that users include a diverse set of machine learning algorithms capable of capturing flexible and complex relationships in the data based on the outcome type. We recommend \citep{phillips2023practical} for more details on specifying appropriate learners.

The parameter \texttt{nn\_module} (line 12) specifies the neural network architecture used for \textit{Riesz learning} \citep{chernozhukov2021automatic,chernuzhukov2022riesznet}. In \texttt{crumble}, Riesz learning is used to estimate conditional density ratios without requiring specification of their functional form; see Section 4 of \citep{liu2024general} or Appendix C for further technical details. In this example, \texttt{sequential\_module()} is the default option and defines a neural network with a single hidden layer. The \texttt{torch} package \citep{Falbel2025torch} provides additional flexibility for customizing the network architecture. To avoid potential installation and runtime issues when using \texttt{torch}, we strongly recommend users to follow the tutorial on the official website (\href{https://torch.mlverse.org/start/}{https://torch.mlverse.org/start/}), which provides a detailed installation guide.

The parameter \texttt{control} (line 13) specifies additional tuning parameters. The parameter \texttt{crossfit\_folds} determines the number of folds used for \textit{cross-fitting} (see Section 4 of \citep{liu2024general} or Appendix C for details), and \texttt{epochs} specifies the number of passes over the full training dataset during neural network training. 

We noticed that researchers need guidance on choosing proper number of folds and epochs. For the number of folds, we recommended that users consider a moderate number of folds (e.g., 5 or 10) in practice. If the number of folds is too small, results may be unstable across repeated runs due to different random seeds, and the estimator may need to satisfy a Donsker class condition, which restricts the complexity of the function class \citep{williams2025re}. Conversely, if the number of folds is too large, computational cost may become a practical concern. The optimal number of training epochs can often be determined using techniques such as early stopping. Although such functionality is not currently implemented in \texttt{crumble}, an empirical approach is to specify a small number of epochs (e.g., 10) and conduct a sensitivity analysis by considering slightly larger or smaller values (e.g., 12 or 8). The chosen value is deemed appropriate if the results remain stable and the substantive conclusions do not change across these settings. Otherwise, it may indicate that the number of epochs is not sufficient to ensure the convergence of the neural network model, and a higher number is needed. 

\section{Extension to Non-binary Exposure using modified treatment policies}
\label{nonbinary}
Another important advantage of \texttt{crumble} is that it allows users to define causal effects for non-binary treatment/exposures using \textit{modified treatment policies} (MTPs, \citep{Diaz12,Haneuse2013,diaz2023lmtp}), or \textit{dynamic interventions that depend on the natural value of treatment} \citep{robins2004effects,young2014identification}. 

An MTP is a user-defined function of treatment value $a$ (which can be non-binary) and covariates $w$, which is useful to describe a lot of interventions of interest. For example, if $a$ is the length of exercise time per day (in minutes), $w$ is the systolic blood pressure (SBP, in mmHg), then, a post-intervention treatment value can be defined as
\begin{align}\label{mtpexample}
    d(a, w) = \begin{cases}
    a - 15 & w \ge 130 \\ 
    a & otherwise
    \end{cases}
\end{align}
which can be interpreted as ``reducing the length of exercise time per day by 15 minutes if the systolic blood pressure is greater or equal to 130 mmHg". We refer readers to \citep{hoffman2023introducing} for more examples and a more comprehensive introduction to MTPs.

The idea of incorporating MTPs into mediation analysis is straightforward. Let $d_1(a, w)$ and $d_0(a, w)$ be two MTPs. Then, the MTP version of a mediation parameter can be obtained by replacing all treatment values $a$ with $d_a$ for $a \in \{0, 1\}$ in the counterfactual outcomes that define the mediation parameter. For example, the MTP version of $\psi_{P_1}$, the effects through path $A \to Y$ that are defined in (\ref{eq:P1toP4}), can be written as

%TC:ignore
\[
\psi_{P_1}^{MTP} = \E(Y(d_1, Z(d_1), M(d_1, Z(d_1))) - Y(d_0, Z(d_1), M(d_1, Z(d_1)))).
\]
%TC:endignore
After incorporating MTPs, one may interpret the mediation effects as the average causal contrast between two hypothetical worlds where some of the treatment values change from $d_1$ to $d_0$, holding other treatment values unchanged. For instance, 
let $d_1(a, w)$ be the $d(a, w)$ in (\ref{mtpexample}), and let $d_0(a, w) = a$, then, $\psi_{P_1}^{MTP}$ can be interpreted as the RTE of the intervention ``reducing the length of exercise time per day by 15 minutes if the systolic blood pressure is great or equal to 130 mmHg" through path $A \to Y$. This effect can also be interpreted as the natural path-specific effect through $A \to Y$ because $\psi_{P_1}^{MTP}$ is identifiable. 

In \texttt{crumble}, the function \texttt{crumble()} sets the input parameters \texttt{d1} and \texttt{d0} to be the so-called ``treatment" and ``control" values that can be non-binary. This design leverages the fact that by defining $d_1(a, w) = 1, d_0(a, w) = 0$, the MTP version of the mediation parameter reduces to the standard mediation parameter for binary exposures. 

We defer the technical introduction of the MTP version of all mediation parameters we reviewed in \sec\ref{ReviewParameter} to Appendix A.

\section{Case Studies}\label{CaseStudy}
In this section, we illustrate the use of \texttt{crumble} by addressing two mediation questions—one with a binary treatment and the other with a non-binary treatment—based on the Job Search Intervention Study (Jobs II) data \citep{vinokur1997mastery}. Jobs II is a randomized field experiment that investigated the efficacy of a job training intervention among unemployed workers. In the study, participants were randomly assigned to treatment and control groups; both groups were exposed to job-search strategies, but the control group relied more heavily on self-directed learning.

We use the \texttt{jobs} dataset from the \texttt{mediation} package \citep{tingley2014mediation}, which is a commonly used illustrative version of the Jobs II data and contains 899 observations and 17 variables. We emphasize that these case studies should be viewed strictly as demonstrations of \texttt{crumble}’s usage, rather than as sources for substantive inference about program efficacy. This is not only because the dataset has been post-processed, but also because the identification assumptions required for the mediation parameters are unlikely to hold exactly (e.g., due to the presence of unmeasured baseline confounders). The fully reproducible R code for all case studies is provided in Appendix B.

\subsection{Causal Effects with Binary Treatment}
\label{CEbinary}
Our first research question is: \textit{What are the average direct, indirect, and path-specific effects of assignment to the treatment group on depressive symptoms measured at the end of the study, with job-search self-efficacy as a mediator?} We address this question using the mediation parameters discussed throughout this tutorial. Specifically, we define $A$ as a binary indicator of whether a participant was assigned to the treatment group; $Z$ as the binary compliance status, indicating whether the participant actually participated the treatment group (not applicable for the NDE and NIE); $M$ as the level of job-search self-efficacy, measured on a scale from 1 to 5; $Y$ as the post-study measure of depressive symptoms based on the Hopkins Symptom Checklist, with values ranging from 1 to 4.909; and $W$ as the set of baseline confounders, including pre-treatment economic hardship, pre-treatment depressive symptoms, age, sex, occupation, education, marital status, and income level.

Table~\ref{Table1} (left panel) displays the results. Nearly all mediation effects are not statistically significant because their $95\%$ confidence intervals (CIs) include $0$. The RTE through $P_4: A \to M \to Y$ ($\psi_{P_4}$) is (marginally) significant, suggesting that job-search self-efficacy may mediate the average treatment effect (ATE). Notably, this mediator is the same one highlighted in the documentation of the \texttt{mediation} package. %We also note that although the NIE is marginally significant, this result---together with the NDE---should be interpreted with caution because the participant’s compliance status $Z$, a well-known post-treatment confounder, was not included in the analysis. 
Although the falsification test of $H_0: R = 0$ fails to reject the null, this does not imply that $Z$ is not a post-treatment confounder. It is important to note that researchers can only make a conclusion about $Z$ when the falsification test rejects the null, and no conclusion can be made without other information if the falsification test fails to reject the null. Here, we do not display the results of NDE and NIE because the complicance status $Z$ is probably an intermediate confounder that makes natural effects unreliable.

%TC:ignore
\begin{table}[h]
% \caption{Illustrative application results. }
\centering
\small
\begin{tabular}[]{l c c}
\toprule
\label{tab:app}
& Estimate & 95\% CI\\
\hline
\multicolumn{3}{c}{RTEs}\\
$P_1$& -0.022& (-0.055, 0.012)\\
$P_2$ & -0.017&  (-0.04, 0.005)\\
$P_3$ & -0.002 &(-0.014, 0.009)\\
$P_4$ & -0.014 & (-0.029, 0.002)\\
$R$ & 0.008 & (-0.024, 0.039)\\
\midrule
\multicolumn{3}{c}{Other Effects}\\
RIDE& -0.022 & (-0.026, -0.018)\\
RIIE& -0.016 & (-0.06, 0.028)\\
%NDE&  -0.041 & (-0.101, 0.019)\\
%NIE& -0.013 & (-0.027, 0.001)\\
\bottomrule
\end{tabular}
\hfill
\begin{tabular}[]{l c c}
\toprule
\label{tab:app}
& Estimate & 95\% CI\\
\hline
\multicolumn{3}{c}{RTEs}\\
$P_1$& 0.013& (0.003, 0.023)\\
$P_2$ & 0.007&  (0.003, 0.012)\\
$P_3$ & 0.011 &(0.009, 0.013)\\
$P_4$ & 0.021 & (0.013, 0.029)\\
$R$ & -0.003 & (-0.007, 0.002)\\
\midrule
\multicolumn{3}{c}{Other Effects}\\
RIDE & 0.027& (0.011, 0.043)\\
RIIE &  0.018& (0.012, 0.024)\\
%NDE&  0.020& (-0.002, 0.041)\\
%NIE& 0.037& (0.031, 0.042)\\
\bottomrule
\end{tabular}
\caption{Results for two case studies. \textbf{Left}: The results for binary treatment case study; \textbf{Right}: The results for non-binary treatment case study. CI: confidence interval. }
\label{Table1}
\end{table}

%TC:endignore
\subsection{Causal Effects with Non-binary Treatment}
Our second research question is: \textit{What is the average direct, indirect and path-specific effect of ``reducing the income of all participants by one level if their income is at least level two" on depressive symptoms measured at the end of the study when job-search self-efficacy is a mediator}? As before, we address this question using all five mediation parameters. The variables $(Z, M, Y)$ are defined in the same way as in \sec~\ref{CEbinary}. In contrast to the first analysis, the treatment variable $A$---which previously belonged to $W$---is now defined as the income level, ranging from 1 to 5. The original binary treatment assignment indicator is instead included as one of the baseline covariates in $W$. This question corresponds to the following MTPs:
\[
d_1(a, w) =
\begin{cases}
a - 1, & a \ge 2, \\
1, & \text{otherwise},
\end{cases};
\qquad
d_0(a, w) = a.
\]

Table~\ref{Table1} (right panel) displays the results. Nearly all estimated mediation effects are positive and statistically significant. This finding is expected: depressive symptoms tend to increase when participants’ income is reduced. We further note that the test $H_0: R = 0$ fails to reject the null. However, the results also show that $\psi_{P_2}$ and $\psi_{P_3}$ are statistically significantly different from zero, indicating that the pathways $A \to Z$, $Z \to M$, and $Z \to Y$ are all active. This implies the presence of intermediate confounding by $Z$. Consequently, we do not display the results of NDE and NIE, either, as they are not reliable in this case study, and researchers should instead rely on other mediation parameters for valid causal interpretation.

%TC:ignore
\section{Summary}\label{Summary}
This tutorial provides a comprehensive introduction to \texttt{crumble}, an R package for modern causal mediation analysis. We first reviewed three types of mediation parameters in Section~\ref{ReviewParameter}, including natural effects (NDE and NIE), randomized effects (RIDE and RIIE), and RTEs. In practice, if researchers believe that there are no measured or unmeasured intermediate confounders, then we recommend using natural effects because of their clear mechanistic interpretation. Otherwise, we recommend using RTEs rather than randomized effects, since RIIE may fail to capture the true causal mechanism \citep{miles2023causal}, and RIDE, while can still measure the true mechanism, only measures the effects through the combination of $P_1$ and $P_2$. Additionally, randomized effects do not decompose the ATE (as also shown in Table~\ref{Table1}), which implies that RIDE and RIIE may be difficult to interpret when researchers are specifically interested in the proportion of the ATE attributable to direct or indirect pathways.

Following the review, we illustrate the use of \texttt{crumble} in Section \ref{sec:demo} by examining the configuration of an example. We then discuss extensions for identifying mediation effects with non-binary exposures via MTPs in Section~\ref{nonbinary}. Together, these developments enable researchers to freely select and estimate mediation parameters aligned with their scientific questions, without being unduly constrained by data complexity.

We hope this tutorial serves as an up-to-date guide for choosing appropriate mediation parameters under different scenarios—particularly with respect to the presence or absence of intermediate confounding by $Z$—and provides a user-friendly explanation of how to apply \texttt{crumble} in practice.

%TC:endignore
%TC:ignore

\section*{Acknowledgements}

We thank Rui Wang at the University of Washington for helpful discussions. Portions of the material in Section \ref{ReviewParameter} and \ref{sec:demo} were motivated by the \href{https://codex.nimahejazi.org/ser2025_mediation_workshop/index.html}{mediation workshop} at the Society for Epidemiologic Research (SER) 2025. Iv\'an D\'iaz and Kara Rudolph were supported through a Patient-Centered Outcomes Research Institute (PCORI) Project Program Funding Award (ME-2021C2-23636-IC) and through the National Institute on Drug Abuse (R01DA053243).

\bibliographystyle{plainnat}
\bibliography{refs}

\newpage
\appendix

\section{Incorporating MTPs into common mediation parameters}
This section briefly introduces the definition and identifications of the common mediation parameters reviewed in \sec 3 when MTPs are incorporated for non-binary treatment/exposures. Details on the estimation of these parameters are provided in \sec 5 of \citep{liu2024general}. In what follows, we denote $d_1 \triangleq d_1(a, w)$ and $d_0  \triangleq d_0(a, w)$ as two user-defined MTPs.
\paragraph{Natural Direct and Indirect Effects (NDE and NIE)} The MTP versions of natural effects are defined as 
\begin{align*}
    \operatorname{NIE} & = \E[Y(d_1, M(d_1)) - Y(d_1, M(d_0))], \\ 
    \operatorname{NDE} & = \E[Y(d_1, M(d_0)) - Y(d_0, M(d_0))].
\end{align*}
Under the same set of assumptions of natural effects (Assumptions A1 to A3 in the main paper), the MTP versions of NDE and NIE are identified as 

\begin{align*}
    \operatorname{NIE}  = \psi^{natural}(1, 1) - \psi^{natural}(1, 0), \operatorname{NDE}  = \psi^{natural}(1, 0) - \psi^{natural}(0, 0),
\end{align*}
where 
\begin{align*}
    \psi^{natural}(a_1, a_2) = \int_{a, w} \int _m \E(Y \mid d_{a_1}(a, w) , m, w) \dd \P (m \mid d_{a_2}(a, w), w) \dd \P(a, w).
\end{align*}
\paragraph{Randomized Direct and Indirect Effects (RIDE and RIIE)} The MTP versions of randomized effects are defined as 
\begin{align*}
    \operatorname{RIIE} = \E[Y(d_1, G(d_1)) - Y(d_1, G(d_0))], \\ 
    \operatorname{RIDE} = \E[Y(d_1, G(d_0)) - Y(d_0, G(d_0))],
\end{align*}
where $G(d_a) \sim M(d_a) \mid W$ for $a \in \{0, 1\}$. Under the same set of assumptions (Assumptions A4 and A5 in the main paper), RIDE and RIIE are identified as 
\begin{align*}
    \operatorname{RIIE} = \psi ^{rand}(1, 1) - \psi ^{rand}(1, 0), \operatorname{RIDE} = \psi ^{rand}(1, 0) - \psi ^{rand}(0, 0),
\end{align*}
where 
{\footnotesize 
\begin{align*}
    \psi^{rand} (a_1, a_2) & = \int_{a, w}\int_m \int_z\E(Y \mid d_{a_1}(a, w), z, m,w) \dd \P( z \mid d_{a_1}(a, w), w) \dd \P (m \mid d_{a_2}(a, w), w) \dd \P (a, w).
\end{align*}}
\paragraph{Recanting-twin Effects (RTEs)} See Appendix A in \citep{liu2024general}. 
\section{Reproducible R code for case studies}

{\small
\begin{lstlisting}[language=R]
library(causalweight)
library(crumble)
library(mlr3superlearner)
library(mlr3extralearners)
library(mediation)

library(dplyr)
library(tidyverse)

set.seed(1234)

jobs <- jobs %>% 
  mutate(treat = as.factor(treat))

jobs <- jobs %>% 
  mutate(income = case_when(
    income == "lt15k" ~ 1,
    income == "15t24k" ~ 2,
    income == "25t39k" ~ 3,
    income == "40t49k" ~ 4,
    TRUE ~ 5
  ))

# Case study with Binary Treatment
crumble(
  data = jobs,
  trt = "treat", 
  outcome = "depress2",
  covar = c("econ_hard", "depress1", "sex", "age", "occp", 
            "marital", "educ", "income"),
  mediators = c("job_seek"),
  d0 = \(data, trt) factor(rep(0, nrow(data)), levels = c("0", "1")), 
  d1 = \(data, trt) factor(rep(1, nrow(data)), levels = c("0", "1")), 
  effect = "N",
  learners = c("mean", "glm", "ranger"), 
  nn_module = sequential_module(),
  control = crumble_control(crossfit_folds = 1L, epochs = 20L)
)

crumble(
  data = jobs,
  trt = "treat", 
  outcome = "depress2",
  covar = c("econ_hard", "depress1", "sex", "age", "occp", 
            "marital", "educ", "income"),
  mediators = c("job_seek"),
  moc = c("comply"),
  d0 = \(data, trt) factor(rep(0, nrow(data)), levels = c("0", "1")), 
  d1 = \(data, trt) factor(rep(1, nrow(data)), levels = c("0", "1")), 
  # The RIDE and RIIE were obtained by setting effect = "RI".
  effect = "RT", 
  learners = c("mean", "glm", "ranger"), 
  nn_module = sequential_module(),
  control = crumble_control(crossfit_folds = 1L, epochs = 20L)
)

# Case study with Non-binary Treatment

crumble(
  data = jobs,
  trt = "income", 
  outcome = "depress2",
  covar = c("econ_hard", "depress1", "sex", "age", "occp", 
            "marital", "educ", "treat"),
  mediators = c("job_seek"),
  moc = c("comply"),
  d0 = \(data, trt) data[[trt]], 
  d1 = \(data, trt) ifelse(data[[trt]] > 1, data[[trt]] - 1, data[[trt]]), 
  effect = "RI",
  learners = c("mean", "glm", "ranger"), 
  nn_module = sequential_module(),
  control = crumble_control(crossfit_folds = 1L, epochs = 20L)
)

crumble(
  data = jobs,
  trt = "income", 
  outcome = "depress2",
  covar = c("econ_hard", "depress1", "sex", "age", "occp", 
            "marital", "educ", "treat"),
  mediators = c("job_seek"),
  d0 = \(data, trt) data[[trt]], 
  d1 = \(data, trt) ifelse(data[[trt]] > 1, data[[trt]] - 1, data[[trt]]), 
  effect = "N",
  learners = c("mean", "glm", "ranger"), 
  nn_module = sequential_module(),
  control = crumble_control(crossfit_folds = 1L, epochs = 20L)
)
\end{lstlisting}}

\section{Technical Details for Estimation: Challenges and Solutions in \texttt{crumble}}

In this section, we briefly introduce the two estimation challenges that come up when either $M$ or $Z$ is continuous or high-dimensional. Then, we explain how \texttt{crumble} addresses these estimation challenges by examining its implementation details. 

\subsection{Challenges in Estimation}
One challenge comes from the necessity to estimate a high-dimensional integral in some identification results of the mediation parameters introduced in \sec \ref{ReviewParameter}; the other one comes from estimating the conditional density ratios that show up in inverse-probability weighted estimators. These challenges may be addressed with the help of modern machine learning tools, but the statistical theory required must go beyond standard parametric modeling. The efficient influence function (EIF) is a central object to address these problems because incorporating it into the estimation process can lead to estimators with valid statistical inference (e.g., confidence intervals) even when data-adaptive methods (e.g., machine learning) are used to estimate nuisance parameters. A detailed discussion of the EIF is beyond the scope of this tutorial, and we refer interested readers to \citep{fisher2021visually,kennedy2022semiparametric,hines2022demystifying,renson2025pulling}. Here, we only introduce notation for the statistical functionals considered in this tutorial. We use $\psi({\F})$ to denote a statistical functional evaluated under a distribution ${\F}$. In particular, $\psi({\F})$ represents the true value when ${\F} = {\P}$ and the estimated value when ${\F} = \hat{\P}$, an estimated distribution. For example, if $\psi$ corresponds to $\psi^{{natural}}$, then $\psi({\P}) \triangleq \psi^{{natural}}(a_1, a_2)$ denotes the parameter under the true distribution $\P$ (i.e., the target estimand), whereas $\psi(\hat{{\P}}) \triangleq \hat{\psi}^{{natural}}(a_1, a_2)$ denotes a plug-in estimate. In addition, we use $\varphi$ to denote the efficient influence function (EIF) of $\psi$, and write $\varphi(X;{\F})$ to represent the EIF evaluated at data $X$ under distribution $\F$.

\subsubsection{Difficulty in Estimating the Plug-in Estimator}\label{estimateplugin}
We note that if a statistical functional $\psi(\P)$ can be written as a repeated conditional expectation of the outcome $Y$ (e.g. $\psi(\P) = \psi^{natural}(a_1, a_2)$), then the \textit{sequential regression} framework, widely used in estimating ATE \citep{naimi2017introduction} or longitudinal causal effects \citep{diaz2021nonparametricmtp}, for example, can be applied to estimate $\psi(\hat \P)$ without worrying about the data complexity. However, not all identification formulas we reviewed in \sec \ref{ReviewParameter} are repeated conditional expectations of $Y$. In fact, some of the identification formulas become hard to estimate when either $M$ or $Z$ is continuous and/or high-dimensional. For example, if $\psi(\P) = \psi^{rand}(a_1, a_2)$, where we recall 
\begin{align*}
    \psi^{rand} (a_1, a_2) & = \int_w\int_m \int_z\E(Y \mid A = a_1, z, m,w) \dd \P( z \mid A = a_1, w) \dd \P (m \mid A= a_2, w) \dd \P (w) \\
    & = \E \left[ \E \left\{ {\color{red} \int_z \E(Y \mid A = a_1, z, M,W) \dd \P( z \mid A = a_1, W)} \mid A = a_2, W \right\} \right] ,
\end{align*}
then, the first step before running regressions to obtain $\psi(\hat \P)$ is to numerically evaluate the integral $\int_z \hat \E(Y \mid A = a_1, z, M) \dd \hat \P( z \mid A = a_1, W)$. When $Z$ is continuous and/or high-dimensional, this complex integral is difficult to evaluate using existing numerical methods. 
\subsubsection{Difficulty in Estimating the EIF} \label{estimateeif}
Almost every EIF of the identification formula in \sec\ref{ReviewParameter} contains hard-to-estimate density ratios. For example, if $\psi(\P) = \psi^{natural}(a_1, a_2)$, then its EIF is 
\begin{align*}
    \varphi(X; \P)& = \frac{\one(A = a_1)}{\P (A = a_1 \mid W)} {\color{red} \frac{\P(M \mid  A = a_2, W)}{\P(M \mid A = a_1, W)}}[Y - \E(Y \mid A, M, W)] \\ 
    & + \frac{\one(A = a_2)}{\P (A = a_2 \mid W)}\left\{  Q(M, W) - \E [Q(M, W) \mid W, A = a_2] \right\} \\ 
    & + E[Q(M, W) \mid W, A = a_2],
\end{align*}
where we denote $Q(M, W) = \E(Y \mid A = a_1, M, W)$. To get $\varphi(X; \hat \P)$, one needs to obtain the estimated density ratio $\frac{\hat \P(M \mid  A = a_2, W)}{\hat \P(M \mid A = a_1, W)}$ by numerical methods, which is very challenging when $M$ is continuous and/or high-dimensional. Moreover, Assumption 1 in \citep{rudolph2024practical} suggests that the conditions required for the Bayes reparameterization approach to achieve desirable statistical properties are more stringent than those needed for the approach implemented in \texttt{crumble} (see Remark 2 in \citep{liu2024general}).

\subsection{Implementation Details in \texttt{crumble}}

The \texttt{crumble()} function in \texttt{crumble} encompasses all the functionalities for estimation. The success in developing only two statistical functionals allows \texttt{crumble} to be both general and code-light. Specifically, the whole process of estimation can be executed by merely seven lines of R code, shown below:

\begin{verbatim}
	cd <- add_zp(cd, moc, control)
	folds <- make_folds(cd@data, control$crossfit_folds, cd@vars@id, cd@vars@Y)
	thetas <- estimate_theta(cd, thetas, folds, params, learners, control)
	alpha_ns <- estimate_phi_n_alpha(cd, folds, params, nn_module, control)
	eif_ns <- calc_eifs(cd, alpha_ns, thetas, eif_n)
	alpha_rs <- estimate_phi_r_alpha(cd, folds, params, nn_module, control)
	eif_rs <- calc_eifs(cd, alpha_rs, thetas, eif_r)
\end{verbatim}
Line 1 applies the function \texttt{add\_zp()} to transform the data $(W_i, A_i, Z_i, M_i, Y_i)_{i = 1}^n$ into an augmented dataset $(W_i, A_i, Z_i^\pi,Z_i, M_i, Y_i)_{i = 1}^n$ used for future steps. The additional column $(Z_i^\pi)_{i = 1}^n$ is obtained by stratified permutation described in \sec 4.3 in \citep{liu2024general}. The introduction of $\Zp$ enables certain integrals---typically difficult to estimate directly---to be rewritten as conditional expectations that can be evaluated using a sequential regression framework.  For example, the introduction of $\Zp$ allows the following identity 
\[\int f(a_1, z, m, w)\dd\P(z\mid a_2, w) = \E[f(a_1,\Zp,M,W)\mid A=a_2, M=m, W=w]\]
to hold, where $\int f(a_1, z, m, w)\dd\P(z\mid a_2, w)$ is hard to estimate when $Z$ is continuous and/or high-dimensional, while $\E[f(a_1,\Zp,M,W)\mid A=a_2, M=m, W=w]$ can be estimated through sequential regression. 

Line 2 uses the function \texttt{make\_folds()} to allow the usage of cross-fitting, which is a strategy to allow using flexible regression techniques to estimate the nuisance functions without imposing a strict Donsker assumption on the class of functions, while maintaining the asymptotic linearity and weak convergence properties. Additional technical details are provided in \sec 4.1 in \citep{liu2024general}.

Line 3 uses the function \texttt{estimate\_theta()} to get an estimate of the mediation parameter that user wants, depending on the input parameter \texttt{effect} in \texttt{crumble()}, which is also the input parameter \texttt{params} in \texttt{estimate\_theta()}. In \citep{liu2024general}, the mediation parameters are categorized into two types: \textit{natural} and \textit{randomized} parameters, because as suggested by Proposition 1, each mediation parameter can be written as either a natural or randomized parameter parametrized by a vector of fixed binary interventions. Depending on the mediation parameter user chooses, \texttt{estimate\_theta()} will run a sequential regression for either {natural} or {randomized} parameter accordingly. 

Next, lines 4-7 return the estimate of the EIF of the mediation parameter selected by the user, which is the EIF of either a natural or randomized parameter. Lines 4-5 are executed when such mediation parameter is a natural parameter, while lines 6-7 are executed otherwise. The EIF can be expressed as as a sum of several components, where each component consists of one or more conditional density ratios multiplied by a residual term. As described in \citep{liu2024general}, the residual terms are estimated using sequential regression, while the conditional density ratios are estimated via \textit{Riesz learning}. Riesz learning is a recently popular framework that helps transforming the problem of estimating a conditional density ratio to an unconstrained optimization problem (Proposition 2 in \citep{liu2024general}). This approach avoids the direct specification or estimation of density functions, since solving the optimization problem does not require explicit knowledge of the density ratio form. Readers are referred to \sec 4.1 and \sec 4.2 in \citep{liu2024general} for more technical details about lines 4-7, and \citep{williams2025riesz} for another tutorial about Riesz regression, which was public after \citep{liu2024general}. 

We also note that \texttt{estimate\_phi\_n\_alpha()} and \texttt{estimate\_phi\_r\_alpha()} in lines 4-7 rely on an input parameter \texttt{nn\_module} because \texttt{crumble} solves the unconstrained optimization problems using deep learning or neural network training \citep{lecun2015deep}. This approach naturally solves optimization problems with respect to functions, and is one of the few machine learning regression procedures with off-the-shelf software (e.g. \texttt{torch} package in R) that allows specification of custom loss functions. Development of alternative optimization algorithms for estimating RRs is left for future work.

\end{document}